%% file: GAOarxiv.tex
\documentclass[12pt]{article}

\usepackage{hyperref}
\usepackage{amssymb}
\usepackage{amsmath}
\usepackage{amsfonts}
\usepackage{graphicx}
\usepackage{epstopdf}
\usepackage{caption}
\usepackage[title]{appendix}
\newtheorem{theorem}{Theorem}[section]
\newtheorem{lemma}{Lemma}[theorem]
\font\myfont=cmr12 at 10pt

\title{\vspace{-5.0cm}\Large Pricing of the Geometric Asian Options Under a Multifactor Stochastic Volatility Model}
\date{\vspace{-5ex}}
 \author{ \myfont Gifty Malhotra \\ \myfont\href{mailto:giftymalhotradtu@gmail.com}{giftymalhotradtu@gmail.com}
\qquad \and \myfont R. Srivastava \\ \myfont \href{mailto:rsrivastava@dce.ac.in}{rsrivastava@dce.ac.in}
\and \myfont H.C. Taneja \\ \myfont \href{mailto:hctaneja@dce.ac.in}{hctaneja@dce.ac.in}\\
\myfont Department of Applied Mathematics, Delhi Technological University,\\
\myfont Delhi(India)-110042}

\begin{document}

\maketitle
\begin{abstract}
\noindent This paper focuses on the pricing of continuous geometric Asian options (GAOs) under a multifactor stochastic volatility model. The model considers fast and slow mean reverting factors of volatility, where slow volatility factor is approximated by a quadratic arc. The asymptotic expansion of the price function is assumed, and the first order price approximation is derived using the perturbation techniques for both floating and fixed strike GAOs. Much simplified pricing formulae for the GAOs are obtained in this multifactor stochastic volatility framework. The zeroth order term in the price approximation is the modified Black-Scholes price for the GAOs. This modified price is expressed in terms of the Black-Scholes price for the GAOs. The accuracy of the approximate option pricing formulae is established, and the model parameter is also estimated by capturing the volatility smiles.
\end{abstract}
\maketitle

\section{Introduction}
\label{sec1}
Asian options, first introduced in $1987$, are the path dependent options such that their payoff depends upon the average of the underlying asset price over the option period. Asian options are much attractive because these are generally less volatile and provide a cheaper way of hedging than their path-independent counterparts. These options are commonly traded in the currency and the commodity markets. The averaging of the underlying asset price in these options can either be arithmetic or geometric. Geometric Asian options (GAOs) are relevant and more attractive than the arithmetic ones because their closed form solutions similar to the Black-Scholes formula are available. GAOs have also been used to approximate the arithmetic Asian option prices (for instance, see~\cite{19}). Mostly the trading of the Asian options is in a discrete sampled way but the daily sampling can be approximated by the continuous sampling case (\cite{23}). Here the continuously sampled GAOs are considered.\\
\indent Asian options have been extensively studied in literature. For the arithmetic Asian options pricing, numerical approximations are given in~\cite{19}, Laplace transform formulae are derived in~\cite{9}, lower and upper bounds are given in~\cite{17}, Monte carlo and Laplace inversion formula is compared in~\cite{7} and a one-dimensional PDE is derived in~\cite{20}. The dimension reduction technique given in~\cite{21} is generalized in~\cite{4} by considering random volatility. The spectral expansions are discussed in~\cite{14}. The pricing based on fourier-cosine expansions is given in \cite{010} etc. For the GAOs, explicit expressions of prices have been derived in~\cite{2,12,23} under the Black-Scholes framework, in \cite{13} under the Heston stochastic volatility model, in \cite{16} using constant elasticity of variance (CEV) model, in \cite{22} using the perturbation techniques in a single factor stochastic model and in \cite{0010} using Barndorff-Nielsen and Shephard model under a stochastic volatility jump model framework, etc.

 The constant volatility assumption of the Black-Scholes framework does not hold well in the actual market scenario. Empirically, the volatility is random which is contrary to the Black-Scholes constant volatility hypothesis. The stochastic volatility models are very well known in the option valuation as they are able to incorporate many features of  volatility namely volatility smile, volatility clustering, mean reversion of volatility, etc. (see~\cite{8}). The single factor stochastic volatility models though generates the volatility smile, but are not able to explain its time varying nature. These models can be extended further to include more factors of stochastic volatility varying on the different time scales. So, the multifactor stochastic volatility models can be think of as the natural extension of these single factor stochastic volatility models. Alizadeh et al.~\cite{1} examined the dynamics of volatility and strongly recommended the two factors of volatility with one highly persistent and another quickly mean reverting factor. Extending this idea Fouque et al.~\cite{5} proposed a two factor stochastic volatility model for the pricing of European options.\\
 \indent Wong and Cheung~\cite{22} considered only fast mean reverting factor of volatility and using asymptotic expansions they derived the approximate pricing formula for continuous GAOs. Malhotra et al.~\cite{15} proposed a multifactor stochastic volatility model by considering both fast and slow mean reverting volatility factors, where the slow factor of volatility is approximated by a quadratic (parabolic) arc, to give the pricing formulae for the European options. Malhotra et al. \cite {015} used this quadratic arc approximation of volatility for the second order moment constraint specifications to calibrate the risk-neutral density function in a two-parametric entropy framework.\\
  \indent In the present work, pricing formulae for the floating and fixed strike continuous GAOs are derived using a multifactor stochastic volatility model. In the GAO setup, there is the path dependence by virtue of geometric averaging of the underlying asset price. Thus, the model used for pricing is an extension of the multifactor stochastic volatility model of Malhotra et al.~\cite{15}, where the pricing formulae were obtained for the path independent options. The accuracy of these formulae is also established. \\
 \indent The rest of the paper is organised as follows: In Section $2$, multifactor stochastic volatility model to be considered for pricing is introduced. Pricing equation and the asymptotic expansion of continuous GAOs is discussed in Section $3$ and $4$ respectively. In Section $5$, the first order approximate option pricing formulae are given. Accuracy of the approximate formulae is discussed in Section $6$. The model parameter is estimated from the market by capturing the European and Asian volatility smiles together in section $7$. Conclusion is given in Section $8$. Detailed mathematical proofs are given in appendices.

 \section{Model Specification}
\label{sec2}
The Asian options are the path-dependent options such that the geometric average of the underlying asset prices over the option period is required for the pricing under the GAO setup.\\
 \indent Let the price of an asset (non-dividend paying) at time $t$ be $X_{t}$. The geometric average of these prices, considered in the interval $[0,t]$, is denoted by $G_{[0,t]}$ and is
expressed as a stochastic process, given below.
\begin{equation}\label{eq4}
 G_{[0,t]} = exp \biggl( \frac{1}{t}\int_{0}^{t}S_{\tau}d\tau \biggr)
\end{equation}
where $S_t = \ln X_t$.
Let $P^{*}$ be the risk neutral probability measure chosen by the market. Under $P^{*}$, the dynamics of stock price is described by
\begin{equation}\label{eq1}
dX_t = rX_t dt + f(Y_t, Z_t) X_t dW_{t}^{x}
 \end{equation}
where $r$ is the risk free rate of interest. $Y_t$ and $Z_t$ are respectively the fast and the slow mean reverting factors of stochastic volatility $f(Y_t, Z_t)$ such that
  \begin{equation}\label{eq2}
dY_t = \frac{1}{\epsilon}(\alpha - Y_t) dt + \frac{\nu\sqrt {2}}{\sqrt{\epsilon}} dW_{t}^{y}
\end{equation}
and
\begin{equation}\label{eq3}
Z_t= \mathcal{P}t^{2}+\mathcal{Q}t+\mathcal{R}+\eta_{t}
\end{equation}
In equation (\ref{eq2}), the fast mean reverting factor $Y_t$ follows an Ornstein-Uhlenbeck (OU) process with the rate of mean reversion $1/\epsilon$, long run mean value $\alpha$ and the volatility parameter $\frac{\nu\sqrt {2}}{\sqrt{\epsilon}}$. It has the long run distribution $N(\alpha,\nu^{2})$ and reverts on a short time scale $\epsilon$.\\
In equation (\ref{eq3}), the approximation of the slow factor of volatility by a quadratic (parabolic) arc is considered. $\eta_{t}$ is the error in the approximation containing the truncation terms and randomness. The value of $\mathcal{P}, \mathcal{Q}$ and $\mathcal{R}$ is specified by the process from which $Z_t$ is approximated. The most commonly used processes to describe the volatility dynamics are the OU process and the Cox-Ingersoll-Ross (CIR) process. Here, the dynamics of $Z_t$ is given by an OU process
\begin{equation}\label{eq5}
dZ_t = k(\alpha^{'} - Z_t) dt + \beta dW_{t}^{z}
\end{equation}
which on simplification is reduced to (\ref{eq3}) (for details refer to \cite{15}) giving
\begin{eqnarray}\label{eq6}
&\nonumber \mathcal{P}=\frac{(Z_{0}-\alpha^{'})}{2}k^{2}\\&
\nonumber \mathcal{Q}=-(Z_0-\alpha^{'})k\\&
\mathcal{R}=Z_{0}
\end{eqnarray}
where $\alpha^{'}$ is the long run mean value about which $Z_t$ is reverting with the rate of mean reversion $k$ and initial value $Z_{0}$ such that $Z_0\neq \alpha^{'}$ to ensure that $\mathcal{P}\neq 0$. $\eta_{t}$ is the error in $Z_t$ involving its volatility parameter $\beta$.   $W_{t}^{x}, W_{t}^{y}$ and $W_{t}^{z}$ are the standard Brownian motions with the correlation structure:
$$E[dW_{t}^{x}.dW_{t}^{y}] = \rho_{xy} dt$$
$$E[dW_{t}^{x}.dW_{t}^{z}] = \rho_{xz} dt$$
$$E[dW_{t}^{y}.dW_{t}^{z}] = \rho_{yz} dt$$
To ensure the positive definiteness of the covariance matrix of the three Brownian motions, the correlation coefficients $\rho_{xy}$, $\rho_{xz}$ and $\rho_{yz}$ satisfy $\rho_{xy}^{2} < 1 , \rho_{xz}^{2} < 1 , \rho_{yz}^{2} < 1$  and $1 + 2\rho_{xy}\rho_{xz}\rho_{yz}-\rho_{xy}^{2}-\rho_{xz}^{2}-\rho_{yz}^{2} > 0$.\\

So, the above system of equations specify the multifactor stochastic volatility model to be considered for the pricing of the GAOs. 
The PDE for the pricing of GAOs is given in the next section.

\section{PDE for the Pricing of Geometric Asian Options}
\label{sec3}

Let $C^{\epsilon}(t,x,y,z,G)$ be the price of a continuous GAO (call) with payoff $h(X_{T},G_{[0,T]})$ or $h(K,G_{[0,T]})$ ( Notation $C^{\epsilon}_{fl}$ and $C^{\epsilon}_{fix}$ will be used for floating and fixed strike GAOs respectively ). Using the Feynmann Kac formula (\cite{6}, pp$48$) , the partial differential equation governing the price comes out to be
\begin{eqnarray}\label{eq7}
        &\nonumber \biggl(\frac{\partial}{\partial{t}}+\frac{1}{2}f^{2}(y,z)x^{2}\frac{\partial^{2}}{\partial x^{2}}+r(x\frac{\partial}{\partial x}- .)+\rho_{xz}\beta f(y,z) x \frac{\partial^{2}}{\partial x\partial z}+\frac{1}{2}\beta^{2}\frac{\partial^{2}}{\partial z^{2}}\\&\nonumber+k(\alpha^{'}-z)\frac{\partial}{\partial z}+\frac{G}{t}\ln(\frac{x}{G})\frac{\partial}{\partial G}+\frac{1}{\epsilon}[(\alpha-y)\frac{\partial}{\partial{y}}+\nu^{2}\frac{\partial^{2}}{\partial{y^2}}]\\&+\frac{1}{\sqrt\epsilon}[\rho_{xy}\nu\sqrt{2}f(y,z) x\frac{\partial^{2}}{\partial x \partial y}+\rho_{yz}\nu\sqrt{2}\beta\frac{\partial^{2}}{\partial y\partial z}]\biggr)C^{\epsilon}=0
        \end{eqnarray}
        with the boundary conditions
\begin{eqnarray}\label{eq8}
&\nonumber C^{\epsilon}_{fl}(T,X_T,Y_T,Z_T,G_{[0,T]})=h(X_T,G_{[0,T]}) = max(X_T - G_{[0,T]} , 0)\\&
           C^{\epsilon}_{fix}(T,X_T,Y_T,Z_T,G_{[0,T]},K)=h(K,G_{[0,T]}) = max (G_{[0,T]} - K , 0)
\end{eqnarray}
for the floating and fixed strike GAOs respectively. $K$ is the strike price.\\
\indent Using the transformation $s=\ln x$ and $u=t\ln\frac{G}{x}$
in (\ref{eq7}) and after solving, we get
\begin{eqnarray*}\label{eq10}
&\nonumber \biggl(\frac{\partial}{\partial{t}}+\frac{1}{2}f^{2}(y,z)\biggl(\frac{\partial}{\partial s}-t\frac{\partial}{\partial u}\biggr)^{2} +\biggl(r-\frac{1}{2}f^{2}(y,z)\biggr)\biggl(\frac{\partial}{\partial s}-t\frac{\partial}{\partial u}\biggr)-r \\&\nonumber+\rho_{sz}\beta f(y,z) \biggl(\frac{\partial}{\partial s}-t\frac{\partial}{\partial u}\biggr)\frac{\partial}{\partial z}
 +\frac{1}{2}\beta^{2}\frac{\partial^{2}}{\partial z^{2}}+k(\alpha^{'}-z)\frac{\partial}{\partial z} +\frac{1}{\epsilon}[(\alpha-y)\frac{\partial}{\partial{y}}+\nu^{2}\frac{\partial^{2}}{\partial{y^2}}]\\& + \frac{1}{\sqrt\epsilon}[\rho_{sy}\nu\sqrt{2}f(y,z) \biggl(\frac{\partial}{\partial s}-t\frac{\partial}{\partial u}\biggr)\frac{\partial}{\partial y}
+\rho_{yz}\nu\sqrt{2}\beta\frac{\partial^{2}}{\partial y\partial z}]\biggr)C^{\epsilon}(t,s,y,z,u)=0
\end{eqnarray*}
It can be written as
\begin{equation}\label{eq11}
\mathcal{L}^{\epsilon}C^{\epsilon}(t,s,y,z,u)=0
\end{equation}
where
\begin{equation*}\label{eq12}
\mathcal{L}^{\epsilon}=\frac{1}{\epsilon}\mathcal{L}_{0}+\frac{1}{\sqrt\epsilon}\mathcal{L}_{1}+\mathcal{L}_{2}
\end{equation*}
such that
\begin{equation}\label{eq13}
\mathcal{L}_{0}=(\alpha-y)\frac{\partial}{\partial{y}}+\nu^{2}\frac{\partial^{2}}{\partial{y^2}},
\end{equation}
\begin{equation}\label{eq14}
\mathcal{L}_{1}=\nu\sqrt{2}\biggl[\rho_{sy}f(y,z) \biggl(\frac{\partial}{\partial s}-t\frac{\partial}{\partial u}\biggr)\frac{\partial}{\partial y }+\rho_{yz}\beta\frac{\partial^{2}}{\partial y\partial z}\biggr],
\end{equation}
and
\begin{eqnarray}\label{eq15}
\mathcal{L}_{2}=&\nonumber \frac{\partial}{\partial{t}}+\frac{1}{2}f^{2}(y,z)\biggl(\frac{\partial}{\partial s}-t\frac{\partial}{\partial u}\biggr)^{2}
+\biggl(r-\frac{1}{2}f^{2}(y,z)\biggr)\biggl(\frac{\partial}{\partial s}-t\frac{\partial}{\partial u}\biggr)-r\\&+\rho_{sz}\beta f(y,z) \biggl(\frac{\partial}{\partial s}-t\frac{\partial}{\partial u}\biggr)\frac{\partial}{\partial z}+\frac{1}{2}\beta^{2}\frac{\partial^{2}}{\partial z^{2}}+k(\alpha^{'}-z)\frac{\partial}{\partial z}.
\end{eqnarray}
Also, the slow factor of volatility $Z_t$ is approximated by a quadratic arc given in (\ref{eq3}).\\
therefore,
\begin{eqnarray*}\label{eq16}
&\nonumber \frac{\partial}{\partial z}=\biggl(\frac{1}{2\mathcal{P}t+\mathcal{Q}+\zeta_t}\biggr)\frac{\partial}{\partial t}\\&
\frac{\partial^{2}}{\partial z^{2}}=\frac{1}{(2\mathcal{P}t+\mathcal{Q}+\zeta_t)^{2}}\biggl[\frac{\partial^{2}}{\partial t^{2}}-\biggl(\frac{2\mathcal{P}+\zeta_{t}^{'}}{2\mathcal{P}t+\mathcal{Q}+\zeta_{t}}\biggr)\frac{\partial}{\partial t}\biggr]
\end{eqnarray*}
where
$$\zeta_t=\frac{\partial \eta_{t}}{\partial t}$$
On substituting this value of $\frac{\partial}{\partial z}$ and $\frac{\partial^{2}}{\partial z^{2}}$ in (\ref{eq14}) and (\ref{eq15}), we get
\begin{equation}\label{eq17}
\mathcal{L}_{1}=\nu\sqrt{2}\biggl[\rho_{sy}f(y,z) \biggl(\frac{\partial}{\partial s}-t\frac{\partial}{\partial u}\biggr)\frac{\partial}{\partial y }+\frac{\rho_{yz}}{2\mathcal{P}t+\mathcal{Q}+\zeta_t}\beta\frac{\partial^{2}}{\partial y\partial t}\biggr],
\end{equation}
and
\begin{eqnarray}\label{eq18}
&\nonumber \mathcal{L}_{2}=
\biggl(1+\frac{k(\alpha^{'}-z)}{2\mathcal{P}t+\mathcal{Q}+\zeta_t}-\frac{1}{2}\beta^{2}\frac{2\mathcal{P}
+\zeta_{t}^{'}}{(2\mathcal{P}t+\mathcal{Q}+\zeta_{t})^{3}}\biggr)\frac{\partial}{\partial{t}}
+\frac{1}{2}f^{2}(y,z)\biggl(\frac{\partial}{\partial s}-t\frac{\partial}{\partial u}\biggr)^{2}\\&\nonumber+\biggl(r-\frac{1}{2}f^{2}(y,z)\biggr)\biggl(\frac{\partial}{\partial s}-t\frac{\partial}{\partial u}\biggr)-r+\frac{\rho_{sz}\beta f(y,z)}{{2\mathcal{P}t+\mathcal{Q}+\zeta_t}} \biggl(\frac{\partial}{\partial s}-t\frac{\partial}{\partial u}\biggr)\frac{\partial}{\partial t}\\&+\frac{1}{2}\beta^{2}\frac{1}{(2\mathcal{P}t+\mathcal{Q}+\zeta_t)^{2}}\frac{\partial^{2}}{\partial t^{2}}
\end{eqnarray}
So, the required pricing equation is (\ref{eq11}) where $\mathcal{L}_{0}, \mathcal{L}_{1}$ and $\mathcal{L}_{2}$ are given by (\ref{eq13}), (\ref{eq17}) and (\ref{eq18}) respectively.

 \section{Asymptotic Price Approximation}
\label{sec4}
Consider the asymptotic expansion of the Asian call option price $C^{\epsilon}$ in the powers of $\sqrt\epsilon$. i.e.
\begin{equation}\label{eq19}
C^{\epsilon}= C_0 + \sqrt\epsilon C_{1} + \epsilon C_{2}+ ...
\end{equation}
(for i = 0,1,2,..., the notation $C_{i}^{fl}$ and $C_{i}^{fix}$ will be used for the floating and fixed strike continuous GAOs respectively).
Put this in the pricing equation (\ref{eq11}) to get
$$\mathcal{L}^{\epsilon}(C_0 + \sqrt\epsilon C_{1} + \epsilon C_{2}+ ...)= 0 $$
i.e.
\begin{equation}\label{eq20}
(\frac{1}{\epsilon}\mathcal{L}_{0}+\frac{1}{\sqrt\epsilon}\mathcal{L}_{1}+\mathcal{L}_{2})(C_0 + \sqrt\epsilon C_{1} + \epsilon C_{2}+ ...)=0
\end{equation}
\indent Equating the terms of various orders equal to zero to get\\
 Terms of order $\frac{1}{\epsilon}$:
\begin{equation}\label{eq21}
 \mathcal{L}_{0}C_{0}=0
 \end{equation}
 Terms of order $\frac{1}{\sqrt\epsilon}$:
 \begin{equation}\label{eq22}
\mathcal{L}_{0}C_{1} + \mathcal{L}_{1}C_{0}=0
 \end{equation}
 Terms of order $1$:
 \begin{equation}\label{eq23}
\mathcal{L}_{0}C_{2} + \mathcal{L}_{1}C_{1} + \mathcal{L}_{2}C_{0}=0
 \end{equation}
 Terms of order $\sqrt \epsilon$:
 \begin{equation}\label{eq24}
 \mathcal{L}_{0}C_{3} +\mathcal{L}_{1}C_{2} + \mathcal{L}_{2}C_{1}=0
 \end{equation}
and so on. Equation (\ref{eq21}), using the expression of $\mathcal{L}_{0}$ given in (\ref{eq13}) implies that the zeroth order term $C_{0}$ in the expansion of $C^{\epsilon}$ is independent of the fast volatility factor $y$ but depends on $z$, the slow factor of volatility.i.e.
\begin{equation}\label{eq25}
C_{0}=C_{0}(t,s,z,u)
\end{equation}
In (\ref{eq22}), $\mathcal{L}_{1}C_{0}=0$ because $C_{0}$ is independent of $y$. Therefore, we are left with
\begin{equation*}\label{eq26}
\mathcal{L}_{0}C_{1}=0
\end{equation*}
which gives that the first order term $C_{1}$ in the price approximation is independent of $y$ but depends on $z$ i.e.
\begin{equation}\label{eq27}
C_{1}=C_{1}(t,s,z,u)
\end{equation}
For the first order price approximation, only the expressions of $C_0$ and $C_1$ are required. Firstly the expression of $C_0$ is obtained. Consider the equation (\ref{eq23}), where $\mathcal{L}_{1}C_{1} = 0$ because $C_{1}$ is independent of $y$ giving
\begin{equation}\label{eq28}
\mathcal{L}_{0}C_{2}+\mathcal{L}_{2}C_{0}=0
\end{equation}
It is a Poisson equation in $C_2$ with respect to $y$ and admits a solution only if the centering condition holds:
\begin{equation*}\label{eq29}
E_{y}[\mathcal{L}_{2}C_{0}] = 0
\end{equation*}
as $C_{0}$ is independent of $y$
\begin{equation}\label{eq30}
\Rightarrow E_{y}[\mathcal{L}_{2}]C_{0} = 0
\end{equation}
Here, $E_{y}[\mathcal{L}_2]$ is the average of operator $\mathcal L_2$ considering the invariant distribution of $y$. The expression of $\mathcal{L}_2$ in (\ref{eq18}) is bit complicated. Therefore, for simplification, neglect the error term in the approximation of the slow volatility factor assuming negligible truncation error and $\beta$ almost zero. Therefore, equation ($\ref{eq18}$) is reduced to the simplified form:
\begin{eqnarray*}\label{eq31}
&\nonumber \mathcal{L}_{2}= \biggl(1+\frac{1-kt+\frac{k^{2}t^{2}}{2}}{1-kt}\biggr)\frac{\partial}{\partial{t}}+\frac{1}{2}f^{2}(y,z)\biggl(\frac{\partial}{\partial s}-t\frac{\partial}{\partial u}\biggr)^{2}\\&\nonumber+\biggl(r-\frac{1}{2}f^{2}(y,z)\biggr)\biggl(\frac{\partial}{\partial s}-t\frac{\partial}{\partial u}\biggr)-r
\end{eqnarray*}
Consider,
\begin{equation}\label{eq32}
\frac{1-kt+\frac{k^{2}t^{2}}{2}}{1-kt}=l
\end{equation}
where, $kt\neq 1$  and $l\neq0$.
This gives
\begin{equation}\label{eq33}
\mathcal{L}_{2}=(1+l)\frac{\partial}{\partial{t}}+\frac{1}{2}f^{2}(y,z)\biggl(\frac{\partial}{\partial s}-t\frac{\partial}{\partial u}\biggr)^{2}+\biggl(r-\frac{1}{2}f^{2}(y,z)\biggr)\biggl(\frac{\partial}{\partial s}-t\frac{\partial}{\partial u}\biggr)-r
\end{equation}
It is worth noticing that
\begin{equation*}\label{eq34}
\mathcal{L}_{2} = \mathcal{L}_{BSA} + l \frac{\partial}{\partial{t}}
\end{equation*}
where $\mathcal{L}_{BSA}$ is the standard Black-Scholes operator for the Asian options.
We call $\mathcal{L}_{2}$ an $l - $ modified Black-scholes operator for the Asian options.
Thus,
\begin{equation}\label{eq35}
E_{y}[\mathcal{L}_{2}]=(1+l)\frac{\partial}{\partial{t}}+\frac{1}{2}\overline \sigma^{2}(z)\biggl(\frac{\partial}{\partial s}-t\frac{\partial}{\partial u}\biggr)^{2}+\biggl(r-\frac{1}{2}\overline\sigma^{2}(z)\biggr)\biggl(\frac{\partial}{\partial s}-t\frac{\partial}{\partial u}\biggr)-r
\end{equation}
is the $l - $ modified Black-Scholes operator for the Asian options with the effective volatility $\overline\sigma(z)$.
The solution of equation (\ref{eq30}) gives the zeroth order term $C_{0}$ of the price approximation and will be called the modified Black-Scholes price for the Asian options.\\ After some calculation, the expression of the modified Black-Scholes price $C_0$ for the Asian options is obtained in terms of the Black-scholes price $B_0$ for the Asian options. It is given (collectively for floating and fixed strike GAOs)by
$$C_{0}^{fl,fix} (t,s,z,u) = \gamma^{fl,fix}(t) B_{0}^{fl,fix} (t,s,\overline\sigma (z),u)$$
or simply,
\begin{equation}\label{eq36}
C_0 (t,s,z,u) = \gamma(t) B_0 (t,s,\overline\sigma (z),u)
\end{equation}
where
\begin{equation}\label{eq37}
\gamma(t) = \biggl[\bigg(\frac{2-kT}{2-kt}\biggr)^{\frac{2}{k}}e^{(T-t)\frac{(2-kt)(2-kT)+2}{(2-kt)(2-KT)}}\biggr]^{M}
\end{equation}
with $kt \neq 2$ for $0<t\leq T$
and
\begin{equation}\label{eq38}
M = \biggl(\frac{1}{B_0}\biggr)\frac{\partial B_0}{\partial t}
\end{equation}
Here, $\frac{\partial B_0}{\partial t}$ is the Black-Scholes theta for the Asian options. $\gamma(t) = 1$ at maturity to satisfy the boundary condition
\begin{equation*}\label{eq39}
C^{fl}_{0}(T,S_T,Z_T,U_T) = h(S_T,U_T)
\end{equation*}
or $$C^{fix}_{0}(T,S_T,Z_T,U_T,K) = h(K,S_T,U_T)$$
After obtaining the expression for $C_0$ we intend to find $C_1$. For this, Consider equation (\ref{eq24}). It is a Poisson equation in $C_3$ with respect to y and admits a solution only if the centering condition holds:
\begin{equation*}\label{eq40}
E_{y}[\mathcal{L}_{2}C_{1}+\mathcal{L}_{1}C_{2}]=0
\end{equation*}
\begin{equation}\label{eq41}
\Rightarrow E_{y}[\mathcal{L}_{2}]C_{1} + E_{y}[\mathcal{L}_{1}C_{2}] = 0
\end{equation}
From equation (\ref{eq28}), $\mathcal{L}_{0}C_{2} = - \mathcal{L}_{2}C_{0}$
$$\Rightarrow \mathcal{L}_{0}C_{2} = -(\mathcal{L}_{2}C_{0}-E_{y}[\mathcal{L}_{2}]C_{0})$$
\begin{equation*}\label{eq42}
\Rightarrow C_{2} = - \mathcal{L}_{0}^{-1}(\mathcal{L}_{2}-E_{y}[\mathcal{L}_{2}])C_{0}
\end{equation*}
putting it in (\ref{eq41}) and on solving we get
\begin{equation*}\label{eq43}
E_{y}[\mathcal{L}_{2}]C_{1} = \mathcal{G}C_0
\end{equation*}
where
\begin{equation}\label{eq44}
\mathcal{G} = V \biggl[ \biggl(\frac{\partial}{\partial s}-t\frac{\partial}{\partial u}\biggr)^{3}-\biggl(\frac{\partial}{\partial s}-t\frac{\partial}{\partial u}\biggr)^{2}\biggr]
\end{equation}
and
\begin{equation}\label{eq45}
V = \frac{\rho_{sy}\nu}{\sqrt 2}E_{y}\biggl[f(y,z) \frac{\partial \phi(y,z)}{\partial y}\biggr]
\end{equation}
$\phi(y,z)$ is the solution of the Poisson equation
\begin{equation}\label{eq46}
\mathcal{L}_{0} \phi (y,z) = f^{2}(y,z) - \overline \sigma ^{2} (z)
\end{equation}
Now, for the floating strike geometric Asian call options,the differentials of the zeroth order term $ C_{0} ^ {fl}$ satisfies
\begin{equation}\label{eq47}
\frac{\partial ^{i+j} C_{0} ^ {fl}}{\partial u^{i} \partial s^{j}} = \frac{\partial ^{i} C_{0}^{fl}}{\partial u^{i}}
\end{equation}
where $i,j = 0,1,2,...$ (see Appendix $A.1$ for the details ). Therefore, for the floating strike GAO, equation (\ref{eq44}) is reduced to
\begin{equation*}\label{eq48}
\mathcal{G}^{fl} = -V \biggl[ t \frac{\partial}{\partial u} - 2t^{2} \frac{\partial ^ 2}{\partial u ^ 2 } + t^{3} \frac{\partial ^ 3}{\partial u ^ 3}\biggr]
\end{equation*}
Similarly, for the fixed strike geometric Asian call option
\begin{equation}\label{eq49}
\frac{\partial ^{i+j} C_{0}^{fix}}{\partial u^{i} \partial s^{j}} = T ^ {j} \frac{\partial ^{i+j} C_{0}^{fix}}{\partial u^{i+j}}
\end{equation}
where $i,j = 0,1,2,...$ (see Appendix $A.1$ for the details ). Thus the corresponding $\mathcal{G}^{fix}$ will be
\begin{equation*}\label{eq50}
\mathcal{G}^{fix} = V \biggl[ (T-t)^{3} \frac{\partial ^ 3}{\partial u ^ 3} - (T-t)^{2} \frac{\partial ^ 2}{\partial u ^ 2 }\biggr]
\end{equation*}
Having the expressions of $\mathcal{G}^{fl}$ and $\mathcal{G}^{fix}$, $C_{1}$ is obtained for the floating and fixed strike GAO using Theorem (\ref{th1}) given below.
\begin{theorem}\label{th1}
Let a function $A(t,s,z,u)$ satisfies
$$E_{y}[\mathcal{L}_{2}]A(t,s,z,u) = \sum_{m =1}^ {n} f_{m}(t)D_{m}(t,s,z,u) $$
with $$A(T,S_T,Z_T,U_T) = 0$$ and $D_{m}(t,s,z,u)$ satisfies the PDE $E_{y}[\mathcal{L}_{2}]D(t,s,z,u)=0$ for every $m = 1,2,...,n$, then A(t,s,z,u) will be of the form
$$A(t,s,z,u)  = -\sum_{m =1}^ {n} \biggl( \int_{t}^{T}\frac {f_{m}(\tau)}{1+l} d\tau \biggr)D_{m}(t,s,z,u)$$
\end{theorem}
where considering $l$ from equation (\ref{eq32}), $1+l = \frac{(2-kt)^2}{2(1-kt)}$. The proof of this theorem is given in Appendix $A.3$.\\
\indent As the $u$-differentials of $C_{0}$ are interchangeable with the modified Black-Scholes operator $E_{y}[\mathcal{L}_{2}]$, so it is clear that the $u$-differentials of $C_{0}^{fl}$ and $C_{0}^{fix}$ will satisfy $E_{y}[\mathcal{L}_{2}]D(t,s,z,u)=0$ . Thus by above theorem,
\begin{equation}\label{eq51}
C_{1}^{fl} = V \biggl[\biggl(\int_{t}^{T} \frac{\tau}{1+l} d\tau\biggr) \frac{\partial}{\partial u} - 2\biggl(\int_{t}^{T} \frac{\tau ^ {2}}{1+l} d\tau\biggr) \frac{\partial ^ 2}{\partial u ^ 2 } + \biggl(\int_{t}^{T} \frac{\tau ^ {3}}{1+l}d\tau\biggr) \frac{\partial ^ 3}{\partial u ^ 3}\biggr]C_{0}^{fl}
\end{equation}
and
\begin{equation}\label{eq52}
C_{1}^{fix} = V \biggl[\biggl(\int_{t}^{T} \frac{(T -\tau)^2}{1+l} d\tau\biggr) \frac{\partial ^ 2}{\partial u ^ 2} - \biggl(\int_{t}^{T} \frac{(T -\tau)^3}{1+l} d\tau\biggr) \frac{\partial ^ 3}{\partial u ^ 3} \biggr]C_{0}^{fix}
\end{equation}
\begin{equation}\label{eq53}
\Rightarrow C_{1}^{fl} = V \biggl[ I_{1} \frac{\partial}{\partial u} -2 I_{2} \frac{\partial ^ 2}{\partial u ^ 2 } + I_{3} \frac{\partial ^ 3}{\partial u ^ 3}\biggr]C_{0}^{fl}
\end{equation}
and
\begin{equation}\label{eq54}
C_{1}^{fix} = V \biggl[ I_{4} \frac{\partial ^ 2}{\partial u ^ 2 } - I_{5} \frac{\partial ^ 3}{\partial u ^ 3}\biggr]C_{0}^{fix}
\end{equation}
where
\begin{eqnarray}\label{eq55}
& \nonumber I_{0} = \frac{-2}{k}\biggl[ k(T-t) (\frac{1}{(2-kT)(2-kt)}) + \ln \frac{2-kT}{2-kt}\biggr]\\&\nonumber
I_{1} = \frac{-2}{k^{2}}\biggl[ k(T-t) (1+ \frac{2}{(2-kT)(2-kt)}) + 3\ln \frac{2-kT}{2-kt}\biggr]\\&\nonumber
I_{2} = \frac{-2}{k^{3}} \biggl[\frac{k^2}{2} (T^{2} - t^{2}) + k(T-t)(3+\frac{4}{(2-kT)(2-kt)}) + 8\ln \frac{2-kT}{2-kt}\biggr]\\&\nonumber
I_{3} = \frac{-2}{k^{4}} \biggl[\frac{k^3}{3} (T^{3} - t^{3}) + \frac{3k^2}{2} (T^{2} - t^{2}) + 8k(T-t)(1+\frac{1}{(2-kT)(2-kt)}) +20\ln \frac{2-kT}{2-kt}\biggr]\\&\nonumber
I_{4} = I_{2} - 2T I_{1} + T^{2} I_{0}\\&
I_{5} = -I_{3} + 3T I_{2} -3T^{2} I_{1} + T^{3}I_{0}
\end{eqnarray}
Clearly, $C_{1}^{fl}$ and $C_{1}^{fix}$ satisfy the boundary condition
\begin{equation*}\label{eq56}
C_{1}(T,S_T,Z_T,U_T) = 0
\end{equation*}
Collectively, for floating and fixed strike GAO
$$C_{1}^{fl,fix} = U^{fl,fix}C_{0}^{fl,fix}$$
The expression of $U$ is given in equations ($\ref{eq53}$) and ($\ref{eq54}$) for the floating and fixed strike GAOs respectively.
\section{First Order Approximated Price}\label{sec5}
The first order approximation to the asymptotic expansion of the GAO call price $C_{fl,fix}^{\epsilon}$ (combined for floating and fixed strike options) is denoted by $\hat C_{fl,fix}^{\epsilon}$ and is given by
\begin{equation}\label{eq57}
C_{fl,fix}^{\epsilon} \approx \hat C_{fl,fix}^{\epsilon}=C_{0}^{fl,fix} + \sqrt {\epsilon} C_{1}^{fl,fix}
\end{equation}
where $C_{0}^{fl,fix}$ is given by equation (\ref{eq36})and $C_{1}^{fl,fix}$ is given by equations (\ref{eq53}) and (\ref{eq54}) for floating and fixed strike options respectively. Unlike \cite{22}, where the approximate price was perturbed around the Black-Scholes price for the GAOs, here the first order approximated price is perturbed around the modified Black-Scholes price $C_{0}^{fl,fix}$ for both floating and fixed strike options. The price is modified by a modification factor $\gamma(t)$ given in (\ref{eq37}). Also, the prices are calculated at the effective volatility $\overline \sigma (z)$ which is estimated by the quadratic arc and not through the standard deviation of the time series of the underlying asset returns.
\section{Accuracy of the Price Approximation}\label{sec6}
The accuracy of the first order price approximation formula (\ref{eq57}) is discussed in this section for the non-smooth pay-off of floating and fixed strike geometric Asian options. The method employed here is on the lines of \cite{05,06} where the accuracy of the first and the second order asymptotics is given for the European vanilla options. The similar method is also employed in \cite{15,22}. The non-smooth pay-off $h$ is regularized by replacing it with modified Black-Scholes formula for the geometric Asian options with time to maturity $\Delta > 0 $ and volatility  $\overline \sigma (z)$ i.e.
\begin{eqnarray}\label{eq58}
\nonumber & h^{\Delta}(s,z,u) = C_{MA}(\Delta ,s,u,\overline \sigma (z))\\ &
 = \gamma (\Delta)B_{A}(\Delta ,s,u,\overline \sigma (z))
\end{eqnarray}
where $C_{MA}$ is a modified Black-Scholes price for the Asian options which is represented in terms of the Black-Scholes price for Asian options $B_{A}$ with modification factor $\gamma (\Delta)$ and volatility $\overline \sigma (z)$. Corresponding to the regularized pay-off $h^{\Delta}$, the regularized option price $C^{\epsilon,\Delta}$ will satisfy
$$\mathcal{L}^{\epsilon}C^{\epsilon,\Delta} = 0$$ where
$$C^{\epsilon,\Delta}(T,s,y,z,u) = h^{\Delta}(s,z,u)$$
and its first order price approximation $\hat C^{\epsilon,\Delta} $ will be
$$\hat C^{\epsilon,\Delta} = C_{0}^{\Delta} + \sqrt{\epsilon} C_{1}^{\Delta} $$
with $$C_{0}^{\Delta}(t,s,z,u) = C_{0}(t-\Delta,s,z,u) = C_{MA}(T-t+\Delta ,s,u,\overline \sigma (z))$$
and $$ C_{1}^{\Delta}(t,s,z,u) = U^{\Delta}C_{0}^{\Delta}(t,s,z,u)$$
\begin{theorem}\label{th2}
Let the volatility function $f(y,z)$ is measurable and bounded away from zero. Then for a fixed $t (< T), s,y,z,u \in \Re$, the accuracy of first order price approximation for geometric Asian option is given by
$$|C^{\epsilon}(t,s,y,z,u)-\hat C^{\epsilon}(t,s,z,u)| \leq b (\epsilon ^ {\frac{1+g}{2}})$$
 for $g<1$ and some constant $b$.
\end{theorem}
The proof of this theorem is a direct implication of three lemmmas given below
\begin{lemma}\label{L1}
Considering a fixed point $(t,s,y,z,u)$ where $t < T$, there exist constants $\Delta_{1} > 0 , \epsilon_{1} > 0$ and $\overline b_{1} > 0 $ so that
$$|C^{\epsilon}(t,s,y,z,u)-C^{\epsilon,\Delta}(t,s,y,z,u)| \leq \overline b_{1}\Delta$$ for all $0<\Delta\leq\Delta_{1}$ and $0<\epsilon\leq\epsilon_{1}$.
\end{lemma}
 It gives that the difference in actual price $C^{\epsilon}$ and the regularized price $C^{\epsilon,\Delta}$ is small. Its proof is given in Appendix $B.1$.
\begin{lemma}\label{L2}
Considering a fixed point $(t,s,y,z,u)$ where $t < T$, there exist constants $\Delta_{2} > 0 , \epsilon_{2} > 0$ and $\overline b_{2} > 0 $ so that
$$|\hat C^{\epsilon}(t,s,z,u)-\hat C^{\epsilon,\Delta}(t,s,z,u)| \leq \overline b_{2}\Delta$$ for all $0<\Delta\leq\Delta_{2}$ and $0<\epsilon\leq\epsilon_{2}$.
\end{lemma}
It gives that the difference in approximated price $C^{\epsilon}$ and its corresponding regularized price $C^{\epsilon,\Delta}$ is small. Its proof is given in Appendix $B.2$.
\begin{lemma}\label{L3}
Considering a fixed point $(t,s,y,z,u)$ where $t < T$, there exist constants $\Delta_{3} > 0 , \epsilon_{3} > 0$ and $\overline b_{3} > 0 $ so that
$$|C^{\epsilon,\Delta}(t,s,y,z,u)-\hat C^{\epsilon,\Delta}(t,s,z,u)| \leq \overline b_{3}\epsilon^{\frac{1+g}{2}}$$ for all $0<\epsilon \leq \epsilon_{3}$, any $g<1$ and uniformly in $\Delta \leq \Delta_{3}$
\end{lemma}
It gives that the difference in regularized price $C^{\epsilon, \Delta}$ and the corresponding regularized first order price approximation $\hat C^{\epsilon,\Delta}$ is small. Its proof is on the lines of lemma $B.3$ of \cite{06} and is given in Appendix $B.3$.\\
Using the above mentioned lemmas \ref {L1}, \ref{L2} and \ref {L3} the proof of theorem \ref{th2} is straightforward. Consider a fixed point $(t,s,y,z,u)$. Take $\overline \epsilon = $ min ($\epsilon_{1},\epsilon_{2},\epsilon_{3}$) and $\overline \Delta = $ min ($\Delta_{1},\Delta_{2},\Delta_{3}$).
\begin{eqnarray}\label{eq59_1}
& \nonumber |C^{\epsilon}-\hat C^{\epsilon}| \leq |C^{\epsilon}-C^{\epsilon,\Delta}| + |C^{\epsilon,\Delta}-\hat C^{\epsilon,\Delta}| + |\hat C^{\epsilon,\Delta}-\hat C^{\epsilon}|\\& \nonumber \leq \overline b_{1} \Delta + \overline b_{2} \Delta + \overline b_{3}(\epsilon ^{\frac{1+g}{2}}) \\& \nonumber \leq 2 max (\overline b_{1},\overline b_{2})\Delta +  \overline b_{3}(\epsilon ^{\frac{1+g}{2}})
\end{eqnarray}
Let $\overline b_{4}= max (\overline b_{1}, \overline b_{2})$ and choose $\Delta=\epsilon$ to get
\begin{eqnarray}\label{eq59}
& \nonumber |C^{\epsilon}-\hat C^{\epsilon}| \leq 2 \overline b_{4}\epsilon +  \overline b_{3}(\epsilon ^{\frac{1+g}{2}}) \\ & \leq b (\epsilon ^{\frac{1+g}{2}})
\end{eqnarray}
This completes the required proof.
\section{Estimation of Model Parameter}\label{sec7} After establishing the accuracy of the approximated pricing formula of GAOs, we intend to estimate $V$, the only parameter required for the implementation of the approximate formulae. The floating strike call and the fixed strike put options always have the positive vegas. Therefore, the volatility smiles can be captured for these options unlike the floating strike put and the fixed strike call options which may have the negative vegas. The European and Asian smiles are captured together on the lines of method given in \cite{22}.
The European smiles with the model under consideration have already been obtained in \cite{15}.\\ Firstly the case of floating strike call options is considered. Assume the asymptotic expansion of market implied volatility $(I)$ for the floating strike options i.e.
$$I^{fl} = I_{0}^{fl} + \sqrt\epsilon I_{1}^{fl} + \epsilon I_{2}^{fl} + ... $$
The implied volatility $I^{fl}$ is obtained by matching the modified Black-Scholes price for Asian options with the price $C_{fl}^{\epsilon}$ of GAO. This gives
$$C_{fl}^{\epsilon}(t,s,z,u,\overline\sigma(z)) = C_{0}^{fl}(t,s,z,u,\sigma = I^{fl})$$
Consider the Taylor expansion of the modified Black-scholes price $C_{0}^{fl}$ about $I_{0}^{fl}$ and the asymptotic expansion of $C_{fl}^{\epsilon}$ given in (\ref{eq19}) to get
$$C_{0}^{fl}(\overline\sigma(z)) + \sqrt\epsilon C_{1}^{fl}(\overline\sigma(z))...  = C_{0}^{fl}(I_{0}^{fl}) + \frac{\partial C_{0}^{fl}(I_{0}^{fl})}{\partial \sigma}\sqrt\epsilon I_{1}^{fl} + ... $$
matching the terms of order zero and $\epsilon$ on both side to get
$$I_{0}^{fl} = \overline\sigma(z)$$
and
$$I_{1}^{fl} = \biggl(\frac{\partial C_{0}^{fl}(I_{0}^{fl})}{\partial \sigma}\biggr)^{-1}C_{1}^{fl}(\overline\sigma(z))$$
The first order approximated value of implied volatility is
$$I^{fl} \approx I_{0}^{fl} + \sqrt\epsilon I_{1}^{fl}$$
For $I_{1}^{fl}$, consider the expression of $C_{1}^{fl}$ from equation (\ref{eq53}) and on rearranging the terms
{\small{
\begin{eqnarray}\label{eq060}
& I^{fl} \approx \overline\sigma(z) + \sqrt\epsilon \biggl[ \biggl(\frac{\partial C_{0}^{fl}(\overline\sigma(z))}{\partial \sigma}\biggr)^{-1} \biggl(a_{fl}\frac{(r\overline\sigma(z))(I_{1} \frac{\partial}{\partial u} -2 I_{2} \frac{\partial ^ 2}{\partial u ^ 2 } + I_{3} \frac{\partial ^ 3}{\partial u ^ 3})C_{0}^{fl}}{\frac{1}{(T-t)}[\frac{1}{k}\log |\frac{kT-2}{kt-2}|+\frac{T-t}{(kT-2)(kt-2)}]}+ d_{fl}\biggr)\biggr]
\end{eqnarray}
\begin{eqnarray}\label{eq061}
\nonumber \Rightarrow \\&(I^{fl} - \overline\sigma(z))\biggl(\frac{\partial C_{0}^{fl}(\overline\sigma(z))}{\partial \sigma}\biggr) \approx a_{fl}^{\epsilon}\biggl(\frac{(r\overline\sigma(z))(I_{1} \frac{\partial}{\partial u} -2 I_{2} \frac{\partial ^ 2}{\partial u ^ 2 } + I_{3} \frac{\partial ^ 3}{\partial u ^ 3})C_{0}^{fl}}{\frac{1}{(T-t)}[\frac{1}{k}\log \frac{2-kT}{2-kt}+\frac{T-t}{(2-kT)(2-kt)}]}\biggr) + d_{fl}^{\epsilon}
\end{eqnarray}}}
where $a_{fl}^{\epsilon} = \sqrt\epsilon a_{fl}$ and $d_{fl}^{\epsilon} = \sqrt\epsilon d_{fl}$. The expressions of $\frac{\partial C^{fl}_{0}}{\partial u}, \frac{\partial ^ 2 C^{fl}_{0}}{\partial u ^ 2 }, \frac{\partial ^ 3 C^{fl}_{0}}{\partial u ^ 3 }$ and $\frac{\partial C^{fl}_{0}}{\partial \sigma}$ are given in Appendix $A.2$. \\
$a_{fl}^{\epsilon}$ and $d_{fl}^{\epsilon}$ are estimated using the simple linear regression and the value of $V$ is obtained from
\begin{equation}\label{eq062}
 V =\frac{ a_{fl} (2r\sigma)}{\frac{2}{(T-t)}[\frac{1}{k}\log \frac{2-kT}{2-kt}+\frac{T-t}{(2-kT)(2-kt)}]}
 \end{equation}
Similarly, for the fixed strike put options
\begin{eqnarray}\label{eq063}
&(I^{fix} - \overline\sigma(z))\biggl(\frac{\partial P_{0}^{fix}(\overline\sigma(z))}{\partial \sigma}\biggr) \approx a_{fix}^{\epsilon}\biggl(\frac{(r\overline\sigma(z))(I_{4} \frac{\partial ^ 2}{\partial u ^ 2 } - I_{5} \frac{\partial ^ 3}{\partial u ^ 3})P_{0}^{fix}}{\frac{1}{(T-t)}[\frac{1}{k}\log \frac{2-kT}{2-kt}+\frac{T-t}{(2-kT)(2-kt)}]}\biggr) + d_{fix}^{\epsilon}
\end{eqnarray}
where $a_{fix}^{\epsilon} = \sqrt\epsilon a_{fix}$ and $d_{fix}^{\epsilon} = \sqrt\epsilon d_{fix}$. The expressions of $\frac{\partial ^ 2 P^{fix}_{0}}{\partial u ^ 2 }, \frac{\partial ^ 3 P^{fix}_{0}}{\partial u ^ 3 }$ and $\frac{\partial P^{fix}_{0}}{\partial \sigma}$ are given in Appendix $A.2$. \\
$a_{fix}^{\epsilon}$ and $d_{fix}^{\epsilon}$ are estimated using the simple linear regression and the value of $V$ is obtained from
\begin{equation}\label{eq064}
 V =\frac{ a_{fix} (2r\sigma)}{\frac{2}{(T-t)}[\frac{1}{k}\log \frac{2-kT}{2-kt}+\frac{T-t}{(2-kT)(2-kt)}]}
 \end{equation}

\begin{figure}[!htb]
\centering
\vspace*{-3cm}
\hspace*{1cm}
\includegraphics[width=18cm]{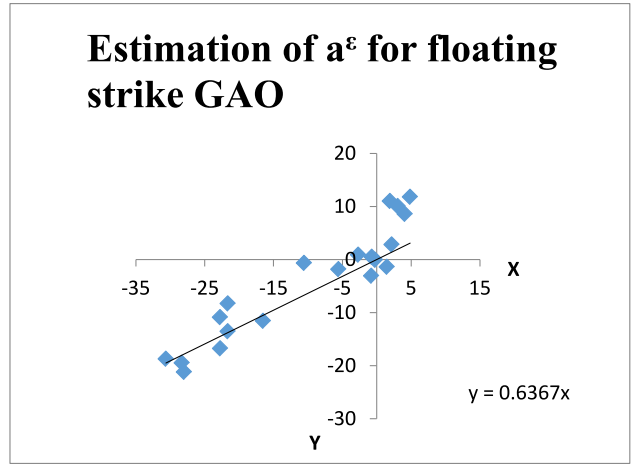}
\vspace{-12.3cm}
\caption{Estimation of $a^{\epsilon}$ from the market data of S $\&$ P 500 index implied volatility for the options with different maturity in the period January 04, 2016 to July 04, 2016.}
\label{est}
\vspace*{-0.2cm}
\end{figure}
{\it{Numerical Illustration}}: The data\footnote{Data sharing is not applicable to this article as no new data were created or analyzed in this study.} of S$\&$P $500$ index options with the floating strike is considered for the period January $4, 2016$ to July $4, 2016$ with the initial stock price $X_{0} = 2013.99$. The other parameters are $k = 2, r = 0.0264, \epsilon = 0.001, Z_{0} = 0.1834, \alpha^{'} = 0.20$. $\overline \sigma (z)$ is estimated from the quadratic arc $\mathcal{P}t^{2}+\mathcal{Q}t+\mathcal{R}$. $T-t$ ranges from $0.01$ to $0.5$.  Firstly, $V$ is calibrated from data. For this, consider equation (\ref{eq061})which can be written as
\begin{equation}
\mathcal{Y} \approx a_{fl}^{\epsilon}\mathcal{X} + d_{fl}^{\epsilon}
\end{equation}
using the simple linear regression with the least square approach, values of $a_{fl}^{\epsilon}$ and $d_{fl}^{\epsilon}$ are estimated from Fig. \ref{est}. The estimated $a^{\epsilon}$ is $0.6367$ and the corresponding $V^{\epsilon}$ is calculated from (\ref{eq062}) which lies in the range $-0.009$ to $-0.003$ for different $T-t$ values.
\begin{figure}[!htb]
\centering
\vspace*{-1.5cm}
\includegraphics[width=18cm]{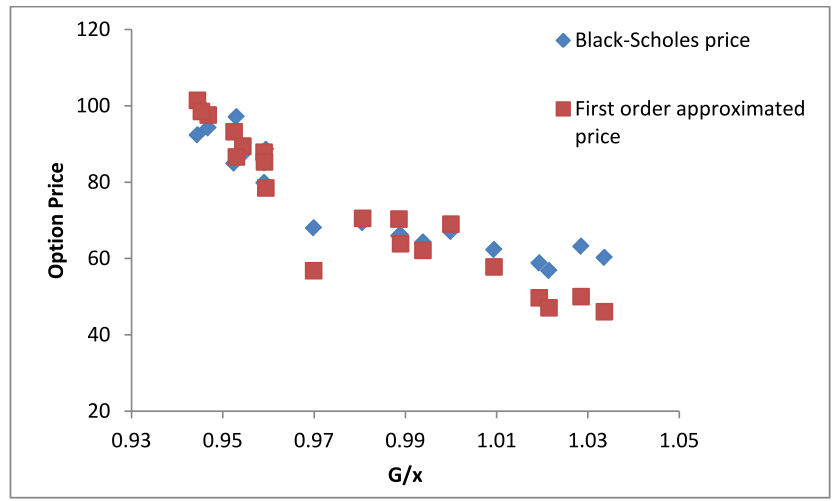}
\vspace{-14cm}
\caption{The comparison of Black-Scholes price of floating strike GAO with the First-order approximated price of floating strike GAO against the moneyness G/x}
\label{op}
\vspace*{-0.2cm}
\end{figure}
With this calibrated $V$ , the first order price approximation of the floating strike GAO is obtained from (\ref{eq57}). The modification factor $\gamma^{fl}$ is in the range $0.7$ to $1.5$ for the different values of $T-t$. This approximated price is compared with the Black-Scholes price for the floating strike Asian options with constant volatility $0.1834$ against the moneyness $G/x$ shown in Fig. \ref{op}.\\
 The effect of modification factor $\gamma$ on the pricing of floating strike GAO is clearly visible in Fig. \ref{op}. For the data set under consideration, first order approximate price differs from the Black-Scholes price and the difference is more prominent for moneyness $G/x$ greater than $1$ giving an improvement over the Black-Scholes prices.
  The calculations get simplified with the quadratic arc approximation of the persistent volatility factor. The formulae obtained are straight forward and easy to implement. $ \it {MATLAB} 2012b$ and $ \it {Excel} 2013$ are used for numerical computations. The run time of the approximated price formulae is also very small and it takes a fraction of second to obtain the results.

\section{Conclusion}
\label{sec:4}
A first order approximated price expression for floating and fixed strike GAO is obtained in a multifactor stochastic volatility framework where slow volatility factor is approximated by a quadratic arc.
The approximated price is perturbed around the modified Black-Scholes price for the Asian options. The accuracy of the approximated formulae is established. There is only one model parameter $V$ which is calibrated by capturing Asian and European smiles together. The first order approximate price for the floating strike GAO is obtained and compared with the corresponding Black-scholes price using the S $\&$ P $500$ index Asian options data. The major advantage of using this model for pricing GAOs is that the calculations get simplified with the quadratic arc approximation of the persistent volatility factor. The pricing formulae thus obtained are straight forward and easy to implement.\\


\begin{appendices}
\section*{Appendix A.1}\label{A1}

The explicit formula of the modified Black-Scholes price $C_0$ for the floating and fixed strike geometric Asian call options is given here in terms of the Black-Scholes formula for the floating and fixed strike geometric Asian call options obtained in \cite{013}. Firstly for the floating strike geometric Asian call,
equation (\ref{eq36}) is $$C_{0}^{fl} = \gamma^{fl}(t) B_{0}^{fl}$$ which gives
\begin{equation}\label{eq60}
C_{0}^{fl} = \gamma^{fl}(t) e^{s}\biggl[ N(d_{1}) - e^{\frac{u}{T}-Q}N(d_{2})\biggr]
\end{equation}
where
\begin{eqnarray}\label{eq61}
&\nonumber d_{1} = \frac{-u + (r+\frac{\overline\sigma ^2}{2}) \frac{T^{2}-t^{2}}{2}}{\overline\sigma \sqrt{\frac{T^{3}-t^{3}}{3}}}\\&\nonumber
d_{2} = d_{1} - \frac{\overline\sigma}{T}\sqrt{\frac{T^{3}-t^{3}}{3}}\\&
Q = (r+\frac{\overline\sigma ^2}{2})\frac{T^{2}-t^{2}}{2T} - \frac{\overline\sigma ^2}{6T^{2}}(T^{3}-t^{3})
\end{eqnarray}
with payoff function
\begin{equation}\label{eq62}
h(S_T,U_T) = e^{S_T} max (1 - e^{\frac{U_T}{T}} , 0)
\end{equation}
$\gamma^{fl}(t)$ is given in (\ref{eq37}).\\

For the fixed strike geometric Asian call, equation (\ref{eq36}) is
$$C_{0}^{fix} = \gamma^{fix}(t) B_{0}^{fix}$$
 \begin{equation}\label{eq63}
C_{0}^{fix} = \gamma^{fix}(t) \biggl[e^{s+\frac{u}{T}-Q} N(\hat d_{1}) - K e^{-r(T-t)}N(\hat d_{2})\biggr]
\end{equation}
where
\begin{eqnarray}\label{eq64}
&\nonumber \hat d_{2} = \frac{\frac{u}{T}+s-\ln K + (r-\frac{\overline\sigma ^2}{2}) \frac{(T-t)^{2}}{2T}}{\frac{\overline\sigma}{T} \sqrt{\frac{(T-t)^{3}}{3}}}\\&
\hat d_{1} = \hat d_{2} + \frac{\overline\sigma}{T}\sqrt{\frac{(T-t)^{3}}{3}}
\end{eqnarray}
with payoff function
\begin{equation}\label{eq65}
h(S_T,U_T) =  max (e^{S_T + \frac{U_T}{T}} - K , 0)
\end{equation}
$\gamma^{fix}(t)$ is given in (\ref{eq37}).
It is clear from (\ref{eq60}) that $\frac{\partial C_{0}^{fl}}{\partial s} = C_{0}^{fl}$ thus $\frac{\partial ^{j} C_{0}^{fl}}{\partial s^{j}} = C_{0}^{fl}$ for every $j = 0,1,2,...$. This results in (\ref{eq47}). \\
Also, for the fixed strike GAOs,
\begin{eqnarray*}
&\nonumber\frac{\partial C_{0}^{fix}}{\partial s} = \frac{\partial}{\partial s}(\gamma^{fix}(t) B_{0}^{fix})
= \gamma^{fix}(t)\frac{\partial B_{0}^{fix}}{\partial s}+B_{0}^{fix}\frac{\partial \gamma^{fix}(t)}{\partial s}\\&
\hspace{-1.2cm}= T\biggl( \gamma^{fix}(t)\frac{\partial B_{0}^{fix}}{\partial u}+B_{0}^{fix}\frac{\partial \gamma^{fix}(t)}{\partial u}\biggr)\\&
\hspace{-5.5cm}= T \frac{\partial C_{0}^{fix}}{\partial u}
\end{eqnarray*}
Thus $\frac{\partial^{j} C_{0}^{fix}}{\partial s^{j}} = T^{j} \frac{\partial^{j} C_{0}^{fix}}{\partial u^{j}}$. This results in (\ref{eq49}).
\section*{Appendix A.2}\label{A2}
For the floating strike call options: The u-differentials and the vega term is given by

$$\frac{\partial C^{fl}_{0}}{\partial u} = - \gamma^{fl}(t) \frac{e^{s+\frac{u}{T}-Q}}{T}N(d_{2})$$
$$\frac{\partial ^ 2 C^{fl}_{0}}{\partial u ^ 2 } = \frac{\gamma^{fl}(t)}{T} \biggl[ \frac{\partial C^{fl}_{0}}{\partial u} + \frac{e^{s+\frac{u}{T}-Q}\Phi(d_{2})}{\overline \sigma \sqrt{\frac{T^{3} - t^{3}}{3}}} \biggr]$$
$$\frac{\partial ^ 3 C^{fl}_{0}}{\partial u ^ 3 } = \frac{\gamma^{fl}(t)}{T} \biggl[ \frac{\partial ^ 2 C^{fl}_{0}}{\partial u ^ 2} + e^{s+\frac{u}{T}-Q}\biggl(\frac{\Phi(d_{2})}{\overline \sigma T \sqrt {\frac{T^{3} - t^{3}}{3}}} + \frac{d_{2}\Phi(d_{2})}{\overline \sigma ^{2} {\frac{T^{3} - t^{3}}{3}}}\biggr) \biggr]$$
and
$$ \frac{\partial C^{fl}_{0}}{\partial \overline \sigma} = \gamma^{fl}(t) e^{s+\frac{u}{T}-Q} \biggl[ \frac{1}{T}\sqrt{\frac{T^{3} - t^{3}}{3}} \Phi(d_{2}) + \frac{\overline\sigma (T-t)^{2} (T+2t)N(d_{2})}{6T^{2}} \biggr]$$

For the fixed strike put options: The u-differentials and the vega term is given by

$$\frac{\partial P^{fix}_{0}}{\partial u} = - \gamma^{fix}(t) \frac{e^{s+\frac{u}{T}-Q}}{T}N(-\hat d_{1})$$
$$\frac{\partial ^ 2 P^{fix}_{0}}{\partial u ^ 2 } = \frac{\gamma^{fix}(t)}{T} \biggl[ \frac{\partial P^{fix}_{0}}{\partial u} + \frac{e^{s+\frac{u}{T}-Q}\Phi(\hat d_{1})}{\overline \sigma \sqrt{\frac{(T-t)^{3}}{3}}} \biggr]$$
$$\frac{\partial ^ 3 P^{fix}_{0}}{\partial u ^ 3 } = \frac{\gamma^{fix}(t)}{T} \biggl[ \frac{\partial ^ 2 P^{fix}_{0}}{\partial u ^ 2} + e^{s+\frac{u}{T}-Q}\biggl(\frac{\Phi(\hat d_{1})}{\overline \sigma T \sqrt {\frac{(T-t)^{3}}{3}}} - \frac{\hat d_{1}\Phi(\hat d_{1})}{\overline \sigma ^{2} {\frac{(T-t)^{3}}{3}}}\biggr) \biggr]$$
and
$$ \frac{\partial P^{fix}_{0}}{\partial \overline \sigma} = \gamma^{fix}(t) e^{s+\frac{u}{T}-Q} \biggl[ \frac{1}{T}\sqrt{\frac{(T-t)^{3}}{3}} \Phi(\hat d_{1}) + \frac{\overline\sigma (T-t)^{2} (T+2t)N(-\hat d_{1})}{6T^{2}} \biggr]$$

where, $N(.)$ and $\Phi(.)$ are respectively the cumulative distribution function and the probability density function for the standard normal variate. \\

\section*{Appendix A.3}\label{A3}
Proof of Theorem (\ref{th1}): This theorem is analogous to the theorem $(3.2)$ of \cite{22}, which has been proved here for the modified Black-scholes operator $E_{y}[\mathcal{L}_{2}]$ which satisfies
$$ E_{y}[\mathcal{L}_{2}]=[1+l]\frac{\partial}{\partial{t}}+\frac{1}{2}\overline\sigma^{2}(z)\biggl(\frac{\partial}
{\partial s}-t\frac{\partial}{\partial u}\biggr)^{2}+\biggl(r-\frac{1}{2}f^{2}(y,z)\biggr)
\biggl(\frac{\partial}{\partial s}-t\frac{\partial}{\partial u}\biggr)-r$$
Here the case $n=1$ is considered only and the proof will be parallel for any other value of $n$. Consider,
$$A(t,s,z,u)  = -\biggl( \int_{t}^{T}\frac {f(\tau)}{1+l} d\tau \biggr)D(t,s,z,u)$$
Clearly, $$A(T,s,z,u) = 0.$$ Also as
$$(1+l)\frac{\partial A}{\partial{t}} = f(t)D(t,s,z,u) - \biggl( \int_{t}^{T}\frac {f(\tau)}{1+l} d\tau \biggr) (1+l)\frac{\partial D}{\partial{t}};$$ and
$$\biggl(\frac{\partial}{\partial s}-t\frac{\partial}{\partial u}\biggr)^{i}A = - \biggl( \int_{t}^{T}\frac {f(\tau)}{1+l} d\tau \biggr)\biggl(\frac{\partial}{\partial s}-t\frac{\partial}{\partial u}\biggr)^{i} D, \forall {i},$$ Therefore,
\begin{eqnarray}
&\nonumber E_{y}[\mathcal{L}_{2}] A(t,s,z,u) = f(t)D(t,s,z,u) - \biggl( \int_{t}^{T}\frac {f(\tau)}{1+l} d\tau \biggr)E_{y}[\mathcal{L}_{2}] D(t,s,z,u)\\&\nonumber \hspace{-2cm} = f(t)D(t,s,z,u)
\end{eqnarray}
because $E_{y}[\mathcal{L}_{2}] D(t,s,z,u) = 0$
This completes the proof of Theorem(\ref{th1})
\section*{Appendix B.1}\label{B1}
Proof of lemma (\ref{L1}): Proof of this lemma is on the lines of lemma $6.1.1$ of \cite{15}. Their technique is similar to \cite{05}.\\ As the payoff functions (\ref{eq62}) and \ref{eq65}) involves both $S_{T}$ and $U_{T}$, Therefore, the joint distribution of $S_{T}$ and $U_{T}$ is required for the risk valuation argument. To reduce the complexity, the joint distribution of $\ln X_{T}$ and $\ln G_{[0,T]}$ is considered in place of the joint distribution of $S_{T}$ and $U_{T}$. For the Simplification, it is considered that $\rho_{sy} = \rho$, $\rho_{yz} = 0$ and $\rho_{sz} = 0$. The proof involves the consideration of the dynamics of a new stochastic process $ \ln {\overline X_{t}}$ which is analogous to the process $\ln X_{t}$ such that
 $$
 d\ln {\overline X_{t}}= \biggl(r-\frac{1}{2}\overline f(t,y,z)^2\biggr)dt+ \overline f(t,y,z)(\rho dW^{y}_{t}+\sqrt{1-\rho^{2}} d\tilde W^{x}_{t})
 $$
 $W^{y}_{t}$ and $\tilde W^{x}_{t}$ are the independent Brownian motions and
\[ \overline f(t,y,z) = \left\{ \begin{array}{ll} f(y,z), & \textnormal{if $t<T$} \\ \overline \sigma (z)   & \textnormal{if $t\geq T$}. \end{array} \right. \]
using the risk neutral valuation under the expectation $E^{*}$,
\begin{eqnarray}
\nonumber & C^{\epsilon}(t,x,y,z,G) - C^{\epsilon,\Delta}(t,x,y,z,G)= E^{*}_{t,x,y,z,G}[e^{-r(T-t)}h(\overline X_{T}, \overline G_{[0,T]})]\\& - E^{*}_{t,x,y,z,G}[e^{-r(T-t+\Delta)}h(\overline X_{T+\Delta}, \overline G_{[0,T+\Delta]})]
\end{eqnarray}
$\Rightarrow$
\begin{eqnarray}
&\nonumber\hspace{-5cm} C^{\epsilon}(t,x,y,z,G)-C^{\epsilon,\Delta}(t,x,y,z,G)= \\&\nonumber E^{*}_{t,x,y,z,G}[E^{*}_{t,x,y,z,G}\{e^{-r(T-t)}h(\overline X_{T},\overline G_{[0,T]})|(W^{y}_{p})_{t\leq p \leq T}\} \\& - E^{*}_{t,x,y,z,G}\{e^{-r(T-t+\Delta)}h(\overline X_{T+\Delta},\overline G_{[0,T+\Delta]} )|(W^{y}_{p})_{t\leq p \leq T}\}]
\end{eqnarray}

The joint distribution of random vector $[\ln {\overline X_{T}}, \ln {\overline G_{[0,T]}}]$ given the path of $(W^{y}_{p})_{t\leq p \leq T}$ is a multivariate normal distribution with a $2$-dimensional mean vector $[\mu_{x}, \mu_{G}]$ and a $2\times 2$covariance matrix
\[\left[ \begin{array}{ll}
   \overline \sigma_{\rho,x}^{2} (T-t)       & \overline \sigma_{\rho,x}^{2} \frac{(T-t)^2}{2T}\\
   \overline \sigma_{\rho,x}^{2} \frac{(T-t)^2}{2T}     & \overline \sigma_{\rho,x}^{2} \frac{(T-t)^3}{3T^2}
\end{array} \right] \]
where,
$$
\mu_{x}=\ln {\overline X_{t}}+ \biggl( r- \frac{\overline \sigma^{2}_{\rho,x}}{2}\biggr)(T-t)+\lambda_{t,T},
$$

$$
\mu_{G}=
\frac{t}{T}\ln {\overline G_{[0,t]}}+\frac{1}{T}\biggl(\ln {\overline X_{t}}(T-t)+ \biggl( r- \frac{\overline \sigma^{2}_{\rho,x}}{2}\biggr)\frac{(T-t)^2}{2}+\int_{t}^{T}\lambda_{t,q}dq\biggr),
$$
$$
\overline \sigma^{2}_{\rho,x}(T-t)= (1-\rho^{2})\int_{t}^{T}\overline f^{2}(p,y,z) dp
$$
 and
$$
\lambda_{t,T}= \rho\int_{t}^{T}\overline f(p,y,z) dW^{y}_{p}- \frac{\rho^{2}}{2}\int_{t}^{T}\overline f^{2}(p,y,z) dp
$$
Similarly, The joint distribution of random vector $[\ln \overline X_{T+\Delta}, \ln \overline G_{[0,T+\Delta]}]$ given the path of $(W^{y}_{p})_{t\leq p \leq T}$ is a multivariate normal distribution with a $2$-dimensional mean vector $[\mu_{x,\Delta}, \mu_{G,\Delta}]$ and a $2\times 2$covariance matrix
\[\left[ \begin{array}{ll}
   \overline \sigma_{\rho,x,\Delta}^{2} (T-t)       & \overline \sigma_{\rho,x,G,\Delta}^{2} \frac{(T-t)^2}{2T}\\
   \overline \sigma_{\rho,x,G,\Delta}^{2} \frac{(T-t)^2}{2T}     & \overline \sigma_{\rho,G,\Delta}^{2} \frac{(T-t)^3}{3T^2}
\end{array} \right] \]
where,
$$
\mu_{x,\Delta}=\ln {\overline {X_{t}}} + \biggl( r- \frac{\overline \sigma^{2}_{\rho,x,\Delta}}{2}\biggr)(T-t) + r\Delta + \lambda_{t,T},
$$

$$
\mu_{G,\Delta}=
\frac{t}{T}\ln {\overline G_{[0,t]}}+\frac{1}{T}\biggl(\ln {\overline X_{t}}(T-t+\Delta)+ \biggl( r- \frac{\overline \sigma^{2}_{\rho,x,G,\Delta}}{2}\biggr)\frac{(T-t)^2}{2}+ r\Delta (T-t + \frac{\Delta}{2}) + \int_{t}^{T}\lambda_{t,q}dq \biggr),
$$
with
$$
\overline \sigma^{2}_{\rho,x,\Delta}(T-t)= \overline \sigma^{2}_{\rho,x}(T-t)+ \overline \sigma^{2}\Delta
$$
$$
\overline \sigma^{2}_{\rho,x,G,\Delta}(T-t)^2 = \overline \sigma^{2}_{\rho,x}(T-t+\Delta)^2 + \rho^{2}\overline \sigma^{2}\Delta^{2}
$$
and
$$
\overline \sigma^{2}_{\rho,G,\Delta}(T-t)^3 = \overline\sigma^{2}_{\rho,x}(T-t+\Delta)^3 + \rho^{2}\overline \sigma^{2}\Delta^{3}
$$
$\gamma(t)$ will always be finite for a finite $t$. Also, using the explicit Black-Scholes formula for the geometric Asian options given in Appendix $A.1$ and under the assumption of bounded $f(y,z)$, it is quite easy to show that
$$
|C^{\epsilon}(t,x,y,z,G) - C^{\epsilon,\Delta}(t,x,y,z,G)| \leq \overline b_{1}\Delta
$$
for some constant $\overline b_{1}$ and some small $\Delta$.

\section*{Appendix B.2}\label{B2}
Proof of this lemma is on the lines of lemma $6.1.2$ of \cite{15}. Their technique is similar to \cite{05}. The proof of the lemma is straightforward. From equations (\ref{eq36}, \ref{eq53}, \ref{eq54}, \ref{eq57}), for the floating and fixed strike geometric Asian call options, the difference
 \begin{eqnarray}
&\nonumber \hat C^{\epsilon}_{fl,fix}(t,s,z,u)-\hat C^{\epsilon,\Delta}_{fl,fix}(t,s,z,u) \\&= \biggl( 1+ G_{1}^{fl,fix}\frac{\partial}{\partial u} + G_{2}^{fl,fix}\frac{\partial^{2}}{\partial u^{2}} + G_{3}^{fl,fix}\frac{\partial^{3}}{\partial u^{3}} \biggr)(C_{0}^{fl,fix} - C_{0}^{fl,fix,\Delta})
 \end{eqnarray}
 where for $i = 1,2,3$ and $j = 0, 1 ,..., 5$, $G_{i}^{fl,fix}$ involves $I_{j}$ and $V$ with their expressions given in (\ref{eq55}) and (\ref{eq45}) respectively. From \cite{05}, $V^{\epsilon} = \sqrt {\epsilon} V$ is bounded. Also for $kt \neq 2$, $\gamma(t)$ , the Black-Scholes price for Asian options and its successive derivatives w.r.t $u$ are differentiable in $t < T$, therefore the same holds for the modified Black-Scholes price for the Asian options. So it can be easily shown that
$$|\hat C^{\epsilon}(t,s,z,u)-\hat C^{\epsilon,\Delta}(t,s,z,u)| \leq \overline b_{2}\Delta$$ for some constant $\overline b_{2}$ and small $\Delta$.
\section*{Appendix B.3}\label{B3}
The proof of this lemma is on the lines of lemma $B.3$ of \cite{06}. All the assumptions given in subsection $(2.1)$ of \cite{06} clearly holds in case of European style GAO where the volatility factors are considered to follow OU process. \\Consider
$C^{\epsilon,\Delta}=C^{\Delta}_{0}+\sqrt\epsilon C^{\Delta}_{1}+ \epsilon C^{\Delta}_{2} + \epsilon\sqrt{\epsilon} C^{\Delta}_{3}- R^{\epsilon,\Delta} $ \\
where $R^{\epsilon,\Delta}$ is the residual for the regularized problem such that
$$\mathcal{L}^{\epsilon}R^{\epsilon,\Delta}=K^{\epsilon,\Delta}$$
with
\begin{eqnarray}\label{eq68}
&\nonumber  K^{\epsilon,\Delta} = \mathcal{L}^{\epsilon} (C^{\Delta}_{0}+\sqrt\epsilon C^{\Delta}_{1}+ \epsilon C^{\Delta}_{2} + \epsilon\sqrt{\epsilon} C^{\Delta}_{3} - C^{\epsilon,\Delta}) \\& = \epsilon(\mathcal{L}_{2}C^{\Delta}_{2}+\mathcal{L}_{1}C^{\Delta}_{3})+ \epsilon\sqrt{\epsilon}(\mathcal{L}_{2}C^{\Delta}_{3})
\end{eqnarray}

Where,
$$ C^{\Delta}_{2} = -\frac{1}{2}\phi(y,z)D_{2}C^{\Delta}_{0}$$
$$C^{\Delta}_{3} = \frac{\mu}{\sqrt 2}\rho \psi_{1}(y,z)D_{1}D_{2}C^{\Delta}_{0} - \frac{1}{2}\phi(y,z)D_{2}C^{\Delta}_{1} $$
$$D_{1} = \biggl(\frac{\partial}{\partial s}-t\frac{\partial}{\partial u}\biggr)$$
$$D_{2} = \biggl(\frac{\partial}{\partial s}-t\frac{\partial}{\partial u}\biggr)^{2}-\biggl(\frac{\partial}{\partial s}-t\frac{\partial}{\partial u}\biggr)$$
and $\psi_{1}(y,z)$ is the solution of Poisson equation
$$\mathcal{L}_{0} \psi (y,z) = f\phi ^{'} - E_{y}[f\phi ^{'}]$$
on putting these values in (\ref{eq68}) and on solving for the floating and fixed strike GAOs, we get
{\small{\begin{eqnarray}
&\nonumber K^{\epsilon,\Delta}_{fl}= \gamma^{fl} \biggl[\epsilon \biggl(\sum_{i=1}^{6}\chi^{fl}_{i,0}(t,T,y,z)\frac{\partial^{i}}{\partial u^{i}}+(T-t)\sum_{i=1}^{6}\chi^{fl}_{i,1}(t,T,y,z)\frac{\partial^{i}}{\partial u^{i}}\biggr)\\& +\epsilon\sqrt{\epsilon}\biggl(\sum_{i=1}^{7}\chi^{fl}_{i,2}(t,T,y,z)\frac{\partial^{i}}{\partial u^{i}} +(T-t)\sum_{i=1}^{7}\chi^{fl}_{i,3}(t,T,y,z)\frac{\partial^{i}}{\partial u^{i}}\biggr)\biggr]B^{\Delta}_{0}
\end{eqnarray}}}
and
{\small{\begin{eqnarray}
&\nonumber K^{\epsilon,\Delta}_ {fix}= \gamma^{fix} \biggl[\epsilon \biggl(\sum_{i=1}^{6}\chi^{fix}_{i,0}(t,T,y,z)\frac{\partial^{i}}{\partial u^{i}}+(T-t)\sum_{i=1}^{2}\chi^{fix}_{i,1}(t,T,y,z)\frac{\partial^{i}}{\partial u^{i}} \\&\nonumber + (T-t)^{2}\sum_{i=1}^{5}\chi^{fix}_{i,2}(t,T,y,z)\frac{\partial^{i}}{\partial u^{i}} +(T-t)^{3}\sum_{i=1}^{6}\chi^{fix}_{i,3}(t,T,y,z)\frac{\partial^{i}}{\partial u^{i}}\biggr) \\&\nonumber +\epsilon\sqrt{\epsilon}\biggl(\sum_{i=1}^{4}\chi^{fl}_{i,4}(t,T,y,z)\frac{\partial^{i}}{\partial u^{i}}+(T-t)\sum_{i=1}^{5}\chi^{fl}_{i,5}(t,T,y,z)\frac{\partial^{i}}{\partial u^{i}} \\&\nonumber + (T-t)^{2}\sum_{i=1}^{5}\chi^{fl}_{i,6}(t,T,y,z)\frac{\partial^{i}}{\partial u^{i}} + (T-t)^{3}\sum_{i=1}^{6}\chi^{fl}_{i,7}(t,T,y,z)\frac{\partial^{i}}{\partial u^{i}} \\& + (T-t)^{4}\sum_{i=1}^{7}\chi^{fl}_{i,8}(t,T,y,z)\frac{\partial^{i}}{\partial u^{i}} + (T-t)^{5}\sum_{i=1}^{7}\chi^{fl}_{i,9}(t,T,y,z)\frac{\partial^{i}}{\partial u^{i}}  \biggr)\biggr]B^{\Delta}_{0}
\end{eqnarray}}}
also at maturity,
\begin{eqnarray}
&\nonumber R^{\epsilon,\Delta}(T) = \epsilon C^{\Delta}_{2}(T,s,y,z,u) + \epsilon\sqrt{\epsilon} C^{\Delta}_{3}(T,s,y,z,u)\\&\nonumber = Q^{\epsilon,\Delta}(s,y,z,u) (say)
\end{eqnarray}
on considering the value of $C^{\Delta}_{2}$ and $C^{\Delta}_{3}$ for both floating and fixed strike GAO at maturity, one can easily show that
$$Q^{\epsilon,\Delta}_{fl,fix}(s,y,z,u) = \gamma^{fl,fix} \biggl[\epsilon \sum_{i=1}^{2}q^{fl,fix}_{i,0}(T,y,z)\frac{\partial^{i}}{\partial u^{i}}+ \epsilon\sqrt{\epsilon}\sum_{i=1}^{3}q^{fl,fix}_{i,1}(T,y,z)\frac{\partial^{i}}{\partial u^{i}}\biggr] B^{\Delta}_{0}$$
Clearly for fixed strike options this value is zero.
Using this terminal condition, residual $R^{\epsilon,\Delta}$ has the probabilistic representation
$$R^{\epsilon,\Delta} = E^{*}_{t,s,y,z,u}\bigg[ e^{-r(T-t)}Q^{\epsilon,\Delta} - \int_{t}^{T}e^{-r(\tau-t)}K^{\epsilon,\Delta}d\tau\biggr]$$
Using lemma $B.4$ of \cite{06}, for both floating and fixed strike options at a fixed $t<T$,
$$\biggl| E^{*}_{t,s,y,z,u}\bigg[ Q^{\epsilon,\Delta}(S_{T},Y_{T},Z_{T},U_{T})]\biggr| \leq b_{3}\epsilon^{\frac{1+g}{2}}$$ and
$$\biggl| E^{*}_{t,s,y,z,u}\bigg[\int_{t}^{T}e^{-r(\tau-t)}K^{\epsilon,\Delta}(s,y,z,u)d\tau\biggr]\biggr| \leq c_{3}\epsilon^{\frac{1+g}{2}}$$
therefore,
$$|C^{\epsilon,\Delta}(t,s,y,z,u)-\hat C^{\epsilon,\Delta}(t,s,z,u)| = |\epsilon C_{2}^{\Delta} + \epsilon \sqrt {\epsilon} C_{3}^{\Delta} - R^{\epsilon, \Delta}|$$
where $C_{2}^{\Delta}$ and $C_{3}^{\Delta}$ are bounded for $t<T$ giving
  $$|C^{\epsilon,\Delta}(t,s,y,z,u)-\hat C^{\epsilon,\Delta}(t,s,z,u)| \leq \overline b_{3}\epsilon^{\frac{1+g}{2}}$$
  for some $\overline b_{3}$ and $g<1$.

\end{appendices}

\pagebreak

\input{referenc}

\end{document}

%% file: referenc.tex
%
%